\documentclass[sn-mathphys-num]{sn-jnl}%

\usepackage{graphicx}%
\usepackage{multirow}%
\usepackage{amsmath,amssymb,amsfonts}%
\usepackage{amsthm}%
\usepackage{mathrsfs}%
\usepackage[title]{appendix}%
\usepackage{xcolor}%
\usepackage{textcomp}%
\usepackage{manyfoot}%
\usepackage{booktabs}%
\usepackage{algorithm}%
\usepackage{algorithmicx}%
\usepackage{algpseudocode}%
\usepackage{listings}%
\usepackage{caption}
\usepackage{booktabs}
\usepackage{graphicx}
\usepackage{tabularray}

\theoremstyle{thmstyleone}%
\newtheorem{theorem}{Theorem}%
\newtheorem{proposition}[theorem]{Proposition}%

\theoremstyle{thmstyletwo}%

\theoremstyle{thmstylethree}%

\usepackage{orcidlink}

\usepackage[braket,qm]{qcircuit}
\usepackage{qcircuit}
\usepackage{comment}
\usepackage{mathtools}

\newcommand{\RR}{{{\mathbb R}}}
\newcommand{\CC}{{{\mathbb C}}}

\newcommand{\Rone}[1]{\textcolor{black}{#1}}
\newcommand{\Rtwo}[1]{\textcolor{black}{#1}}

\usepackage[algo2e]{algorithm2e}
\usepackage{nicematrix}
\usepackage{natbib}
\usepackage{graphicx}
\usepackage{booktabs}
\usepackage{amsmath}
\usepackage{mathtools}
\usepackage{makecell}
\usepackage{tabularx}
\usepackage{fancyvrb,xcolor}
\usepackage{siunitx}

\begin{document}

\title[Article Title]{Quantum Visual Feature Encoding Revisited}

\author*[1,4]{\fnm{Xuan-Bac} \sur{Nguyen}}\email{xnguyen@uark.edu}

\author[1,4]{\fnm{Hoang-Quan} \sur{Nguyen}}\email{hn016@uark.edu}

\author[2,4]{\fnm{Hugh} \sur{Churchill}}\email{hchurch@uark.edu}

\author[3]{\fnm{Samee U.} \sur{Khan}}\email{skhan@ece.msstate.edu}

\author[1,4]{\fnm{Khoa} \sur{Luu}}\email{khoaluu@uark.edu}

\affil*[1]{\orgdiv{Dept. of Electrical Engineering and Computer Science}, \orgname{University of Arkansas}, \orgaddress{\city{Fayetteville}, \postcode{72703}, \state{Arkansas}, \country{USA}}}

\affil[2]{\orgdiv{Dept. of Physics}, \orgname{University of Arkansas}, \orgaddress{\city{Fayetteville}, \postcode{72703}, \state{Arkansas}, \country{USA}}}

\affil[3]{\orgdiv{Dept. of Electrical \& Computer Engineering }, \orgname{Mississippi State University}, \orgaddress{\city{Starkville}, \postcode{39762}, \state{Mississippi}, \country{USA}}}
\affil[4]{\orgdiv{MonArk NSF Quantum Foundry}}

\abstract{Although quantum machine learning has been introduced for a while, its applications in computer vision are still limited. This paper, therefore, revisits the quantum visual encoding strategies, the initial step in quantum machine learning. Investigating the root cause, we uncover that the existing quantum encoding design fails to ensure information preservation of the visual features after the encoding process, thus complicating the learning process of the quantum machine learning models. In particular, the problem, termed the "Quantum Information Gap" (QIG), leads to an information gap between classical and corresponding quantum features. We provide theoretical proof and practical examples with visualization for that found and underscore the significance of QIG, as it directly impacts the performance of quantum machine learning algorithms. To tackle this challenge, we introduce a simple but efficient new loss function named Quantum Information Preserving (QIP) to minimize this gap, resulting in enhanced performance of quantum machine learning algorithms. Extensive experiments validate the effectiveness of our approach, showcasing superior performance compared to current methodologies and consistently achieving state-of-the-art results in quantum modeling.}


\maketitle

\section{Introduction}

Quantum machine learning, as highlighted in \cite{biamonte2017quantum,schuld2015introduction,ciliberto2018quantum,lloyd2013quantum}, represents a promising research direction at the intersection of quantum computing and artificial intelligence. Within this realm, the utilization of quantum computers promises to significantly boost machine learning algorithms by leveraging their innate parallel attributes, thereby showcasing quantum advantages that surpass classical algorithms, as suggested by \cite{harrow2017quantum}.
Due to the substantial collaborative endeavors of academia and industry, contemporary quantum devices, often referred to as noisy intermediate-scale quantum (NISQ) devices \cite{preskill2018quantum}, are now capable of demonstrating quantum advantages in specific meticulously crafted tasks \cite{arute2019quantum,zhong2020quantum}.
An emerging research focus lies in leveraging near-term quantum devices for practical machine learning applications, with a prominent approach being hybrid quantum-classical algorithms \cite{bharti2022noisy,cerezo2021variational}, also referred to as variational quantum algorithms. These algorithms typically employ a classical optimizer to refine quantum neural networks (QNNs), allocating complex tasks to quantum computers while assigning simpler ones to classical computers. 
In typical quantum machine learning scenarios, a quantum circuit utilized in variational quantum algorithms is commonly divided into two components: a data encoding circuit and a QNN. 
On the one hand, enhancing these algorithms' efficacy in handling practical tasks involves the development of various QNN architectures. 
Numerous architectures, including strongly entangling circuit architectures \cite{schuld2020circuit}, tree-tensor networks \cite{grant2018hierarchical}, quantum convolutional neural networks \cite{cong2019quantum}, and even automatically searched architectures \cite{ostaszewski2021reinforcement,ostaszewski2021structure,zhang2022differentiable,du2020quantum}, have been proposed. 
On the other hand, careful design of the encoding circuit is crucial, as it can significantly impact the generalization performance of these algorithms.

Encoding classical information into quantum data is a crucial step, as it directly impacts the performance of quantum machine learning algorithms. 
These algorithms are designed to optimize objective functions, such as classification, using encoded data. 
However, quantum encoding poses significant challenges, especially on near-term quantum devices, as highlighted in previous research \cite{biamonte2017quantum}. 
While phase and amplitude encoding are foundational approaches, recent advancements have popularized parameterized quantum circuits (PQCs) as the most practical strategy for encoding on NISQ devices \cite{benedetti2019parameterized}. 
Nevertheless, despite the prevalence of PQCs, it is essential to utilize the basic encoding methods at the first step, such as phase and amplitude encoding. 
An important question arises regarding whether these encoding strategies guarantee preserving fundamental properties or characteristics of classical data in its quantum form. 

\begin{figure}[t]
    \centering
    \includegraphics[width=0.9\linewidth]{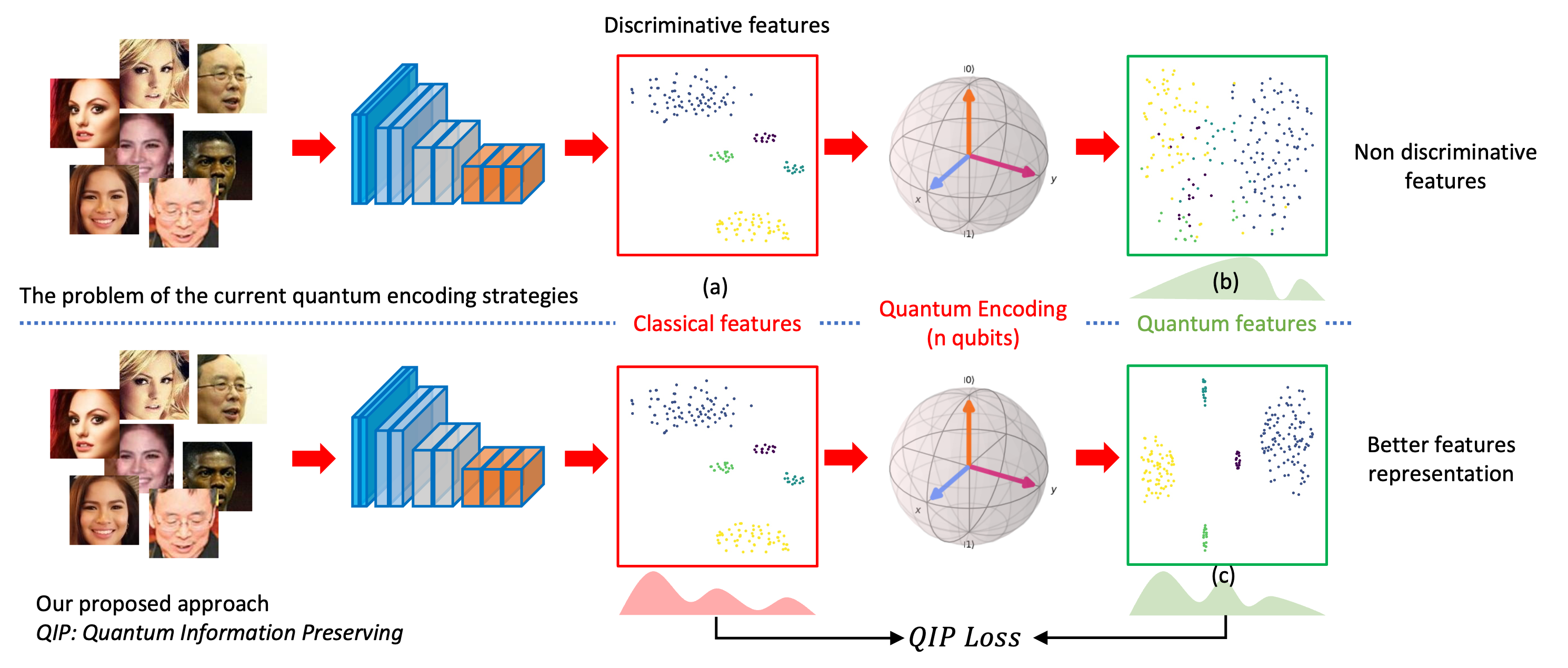}
    \caption{Limitations in current quantum encoding strategies, which result in non-robust feature representations in the quantum feature space and our proposed QIP solution. Figure (b) showcases encoded quantum features. Figure (c) presents our proposed method for enhancing the discriminative of quantum features. 
    }
    \label{fig:problem_introduce}
\end{figure}

\textbf{Contributions of this Work:} This paper has three key contributions. First, we identify a challenge with current visual encoding strategies regarding the preservation of information during the transition from classical to quantum data. Specifically, we observe distinct characteristics between feature spaces in quantum computing compared to their classical counterparts, resulting in lower performance of quantum machine learning algorithms than expected. Second, we introduce a simple but efficient novel training approach to generate classical features conducive to quantum machines post-encoding. This method holds promise for substantially enhancing quantum machine learning algorithms. Finally, our empirical experiments demonstrate the state-of-the-art performance of quantum machine learning across diverse benchmarks.

\section{Related Work}

\subsection{Quantum Computer Vision}

Several quantum techniques are available for computer vision tasks, such as recognition and classification \cite{o2018nonnegative,cavallaro2020approaching}, object tracking \cite{li2020quantum}, transformation estimation \cite{golyanik2020quantum}, shape alignment and matching \cite{noormandipour2022matching,benkner2021q,benkner2020adiabatic}, permutation synchronization \cite{birdal2021quantum}, visual clustering \cite{nguyen2023quantum}, and motion segmentation \cite{arrigoni2022quantum}.
Via Adiabatic Quantum Computing (AQC), O’Malley et al. \cite{o2018nonnegative} applied binary matrix factorization to extract features of facial images. In contrast, Li et al. \cite{li2020quantum} reduced redundant detections in multi-object detection.
Dendukuri et al. \cite{dendukuri2018image} presented the image representation using quantum information to reduce the computational resources of classical computers.
Cavallaro et al. \cite{cavallaro2020approaching} presented multi-spectral image classification using quantum SVM.
Golyanik et al. \cite{golyanik2020quantum} introduced correspondence problems for point sets using AQC to align the rotations between pairs of point sets.
Meanwhile, \cite{noormandipour2022matching} proposed a parameterized quantum circuit learning method for the point set matching problem.
Using AQC to solve the formulated Quadratic Unconstrained Binary Optimization (QUBO), Nguyen et al. \cite{nguyen2023quantum} proposed an unsupervised visual clustering method optimizing the distances between clusters. In contrast, Arrigoni et al. \cite{arrigoni2022quantum} optimized the matching motions of key points between consecutive frames.

\subsection{Hybrid Classical-Quantum Machine Learning}
 
Date et al. \cite{date2020classical} implemented a classical high-performance computing model with an Adiabatic Quantum Processor for a classification task on the MNIST dataset.
Their experiment evaluated two classification models, i.e., the Deep Belief Network (DBN) and the Restricted Boltzmann Machines (RBM).
It is shown that classical computing performs heavy matrix computations efficiently. At the same time, the sampling task is more convenient to quantum computing, as quantum mechanical processes are used to generate samples, making them truly random.
Barkoutsos et al. \cite{barkoutsos2020improving} introduced an improved platform for combinatorial optimization problems using hybrid classical-quantum variational circuits.
It was empirically shown that this approach leads to faster convergence to better solutions for all combinatorial optimization problems on both classical simulation and quantum hardware.
Romero et al. \cite{romero2021variational} presented generative modeling of continuous probability distributions via a Hybrid Quantum-Classical model.
Inspired by convolutional neural networks, Liu et al. \cite{liu2021hybrid} proposed a hybrid quantum-classical convolutional neural network using the quantum advantage to enhance the feature mapping process, the most computationally intensive part of the convolutional neural networks. 
The feature map extracted by a parametrized quantum circuit can detect the correlations of neighboring data points in a complexly large space.

\section{Background}
\subsection{Quantum Basics}

This section provides a concise introduction to fundamental concepts in quantum computing essential for this paper. For a detailed comprehensive review, we refer to \cite{nielsen2001quantum}. In quantum computing, quantum information is typically expressed through $n$-qubit (pure) quantum states within the Hilbert space $\CC^{2^n}$. Specifically, a pure quantum state can be denoted by a unit vector $\ket{\psi} \in \CC^{2^n}$ (or $\bra{\psi}$), where the \textit{ket} notation $\ket{}$ signifies a column vector, and the \textit{bra} notation $\bra{\psi}=\ket{\psi}^\mathsf{T}$ with $\mathsf{T}$ indicating the conjugate transpose, represents a row vector.

Mathematically, the evaluation of a pure quantum state $\ket{\psi}$ is delineated by employing a quantum circuit, often called a quantum gate. It is represented as $\ket{\psi^\prime}=U\ket{\psi}$, where $U$ denotes the unitary operator (matrix) signifying the quantum circuit, and $\ket{\psi^\prime}$ represents the quantum state after the evolution. Standard single-qubit quantum gates encompass the 
Pauli operators.
\begin{align}
    X  := \begin{bmatrix} 0 & 1\\ 1 & 0\end{bmatrix}, 
    Y := \begin{bmatrix} 0 & -i\\ i & 0\end{bmatrix}, 
    Z := \begin{bmatrix} 1 & 0\\ 0 & -1\end{bmatrix},
\end{align}
The corresponding rotation gates denoted by $R_P(\theta)= \text{exp}(-i\theta P/2)$ $ = \cos\frac{\theta}{2} I -i\sin\frac{\theta}{2}P$, where the rotation angle $\theta\in[0,2\pi)$ and $P\in\{X,Y,Z\}$ indicating rotation around $X, Y, Z$ coordinates. In this paper, multiple-qubit quantum gates mainly include the identity gate $I$, the CNOT gate, and the tensor product of single-qubit gates, e.g., $Z\otimes Z$, $Z\otimes I$, $Z^{\otimes n}$ and so on.

Quantum measurement is a method for extracting classical information from a quantum state. 
For example, given a quantum state $\ket{\psi}$ and an observable $H$, one can design quantum measurements to obtain the information $\bra{\psi} H \ket{\psi}$. 
This study concentrates on hardware-efficient Pauli measurements, where $H$ is set as Pauli operators or their tensor products. 
For instance, one might choose $Z_1 = Z\otimes I^{\otimes (n-1)}$, $X_2 = I\otimes X\otimes I^{\otimes (n-2)}$, $Z_1Z_2 = Z\otimes Z\otimes I^{\otimes (n-2)}$, etc., with a total of $n$ qubits.

\subsection{Limitations in Current Quantum Encoding Methods}

Let $\mathbf{v} \in \RR^{d}$ be a typical $d-$dimension vector of a classical computer. We denote $\mathcal{E}(\mathbf{v})$ to be a quantum encoding function that transforms the vector $\mathbf{v}$ into the vector $\ket{\psi} \in \CC^{2^n}$ of quantum states over Hilbert space, where $n$ is the number of qubits.
\begin{equation}
    \ket{\psi} = \mathcal{E}(\mathbf{v})
\end{equation}
    
Specifically, the $\mathcal{E}$ can be amplitude, phase encoding, or PQC. 
It is important to note that the $\ket{\psi}$ represents the qubits' states; for further usage of the quantum machine learning function, it is necessary to extract information from these quantum states. 
To accomplish it, the \textit{observable} denoted as $\mathcal{O}(\ket{\psi})$ is utilized. 
In particular, the observable $\mathcal{O}$ measures the state of every single qubit. Let $\mathbf{q} = [q(0),\dots,q(i),\dots,q({n-1})] \in \RR^{n}$ be a vector of information measured by $\mathcal{O}$ where $q(i)$ is the measurement of $i^{th}$ qubit and formulated as in Eqn. (\ref{eq:qubit_measurement}).
\begin{equation}
    \label{eq:qubit_measurement}
    q(i) = \bra{\psi}\mathcal{O}_i\ket{\psi}
\end{equation}
In the equation above, a different observable $\mathcal{O}_i$ is applied for each qubit. In particular, $\mathcal{O}_i$ is a unitary operator represented by a matrix. Let $P$ be a Pauli operation where $P \in \{X, Y, Z\}$, the $\mathcal{O}_i$ can be further derived as Eqn. (\ref{eq:Oi}).
\begin{equation}
    \label{eq:Oi}
    \mathcal{O}_i = I^{\otimes i} \otimes P \otimes I^{\otimes (n-i-1)}
\end{equation}
According to Eqn. (\ref{eq:Oi}), we can measure the state of a qubit in any coordinates ($X$, $Y$, or $Z$) of the Hilbert space. 

\begin{figure}[t]
    \centering
    \includegraphics[width=0.99\linewidth]{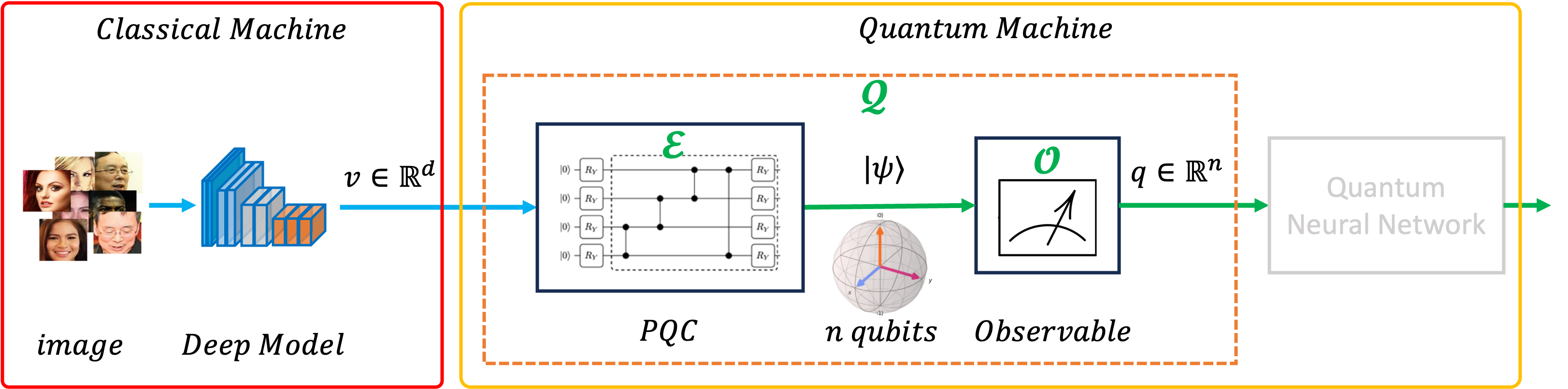}
    \caption{Overview of the hybrid quantum system. The \textcolor{red}{red components} run in the classical machine. The \textcolor{yellow}{yellow box} includes components running on the quantum machine. The dashed \textcolor{orange}{orange box} indicates our focus on this paper. \textbf{Best viewed in color}.}.
    \label{fig:full_pipeline}
\end{figure}

In summary, the relation between quantum information vector $\mathbf{q}$ and classical information vector $\mathbf{v}$ is represented as Eqn. (\ref{eq:full_pipeline}).
\begin{equation}
    \label{eq:full_pipeline}
    \mathbf{v} \in \RR^{d} \xmapsto[\text{Quantum encoding}]{\mathcal{E}(\mathbf{v})} \ket{\psi} \in \RR^{2^n} = \RR^{d} \xmapsto[\text{Measurement}]{\mathcal{O}(\ket{\psi})} \mathbf{q} \in \RR^{n}
\end{equation}
Mathematically, we can define $\mathcal{Q}$ as the function to map $\mathbf{v} \rightarrow \mathbf{q}$ as Eqn. (\ref{eq:function_v_to_q}).
\begin{equation}
    \label{eq:function_v_to_q}
    \mathbf{q} = \mathcal{Q}(\mathbf{v}, \mathcal{E}, \mathcal{O})
\end{equation}
The details of the proposed framework are demonstrated in Fig. \ref{fig:full_pipeline}.
\\
\begin{proposition}\label{prop:qig}
Consider two different quantum state vectors, denoted as $|\psi_1\rangle$ and $|\psi_2\rangle$, and these corresponding quantum information vectors $\mathbf{q}_1$ and $\mathbf{q}_2$. We have $\langle\psi_1|\psi_2\rangle \neq \mathbf{q}_1^\mathsf{T} \mathbf{q}_2$ for any Pauli observable and quantum encoding strategies. 
\end{proposition}

\begin{proof}
As $q(i) = \bra{\psi}\mathcal{O}_i\ket{\psi}$, we have:
\begin{equation}
\begin{split}
    \mathbf{q}_1^\mathsf{T} \mathbf{q}_2 &= \sum_{i=1}^n q_1(i)q_2(i) \\
                       &= \sum_{i=1}^n \langle\psi_1| \mathcal{O}_i |\psi_1\rangle \langle\psi_2| \mathcal{O}_i |\psi_2\rangle \\
                       &= \langle\psi_1| \left( \sum_{i=1}^n \mathcal{O}_i |\psi_1\rangle \langle\psi_2| \mathcal{O}_i \right) |\psi_2\rangle, \\
                       & = \langle\psi_1| A |\psi_2\rangle,
\end{split}
\end{equation}
where $A = \sum_{i=1}^n \left( \mathcal{O}_i |\psi_1\rangle \langle\psi_2| \mathcal{O}_i \right)$. We have to prove that $A \neq I$. That is true because:

\begin{equation}
\label{eq:trace_A}
\begin{split}
    \text{tr}(A)   &= \sum_{i=1}^n \text{tr}(\mathcal{O}_i |\psi_1\rangle \langle\psi_2| \mathcal{O}_i) \\
            &= \sum_{i=1}^n \text{tr}(\mathcal{O}_i \mathcal{O}_i |\psi_1\rangle \langle\psi_2|) \\
            &= \sum_{i=1}^n \text{tr}(|\psi_1\rangle \langle\psi_2|)
\end{split}
\end{equation}
From Proposition \ref{prop:qig}, since $|\psi_1\rangle \neq |\psi_2\rangle$, then $\langle\psi_1|\psi_2\rangle = \text{tr}(|\psi_1\rangle\langle\psi_2|) < 1$. For that reason, we have $\text{tr}(A) < n$ then $A \neq I$ since $\text{tr}(I) = n$. The proposition \ref{prop:qig} has been proven. \Rone{This proposition indicates that no Pauli observable and quantum encoding strategies keep the information when we transform the classical features into quantum features.}
\end{proof}

\subsection{Theoretical Analysis and Problem Visualization}
In this section, we first pre-define the definition of the term \textit{information} as the correlation between pairwise vectors. \\
\textbf{\textit{Theoretical Analysis}}. The goal of encoding $\mathcal{E}$ is to transform a classical feature $\mathbf{v} \in \mathbb{R}^d$ into a quantum state $\ket{\psi} \in \mathbb{R}^d$ using fewer bits while retaining maximum information as much as in the classical one. Assuming $\mathbf{v}$ is a normalized vector and $\mathcal{E}$ represents an amplitude encoding, the preservation of information is evident as $\mathbf{v} = \ket{\psi}$. Additionally, since $\mathcal{E}$ requires fewer than $d$ qubits ($n < d$), it appears to be the optimal choice given these constraints.

However, the limitation of amplitude encoding is its potential unsuitability for many problems. To address this problem, Parametrized Quantum Circuits (PQC) have recently become the most prevalent encoding strategy. PQC incorporates trainable parameters that can be optimized during training, reducing dependencies on specific problems. However, information is not guaranteed to be preserved when representing features in Hilbert spaces of $\ket{\psi}$. Additionally, Proposition \ref{prop:qig} suggests that no observables guarantee uniform discriminability between the features $\ket{\psi}$ and $\mathbf{q}$. Considering these factors, current encoding strategies fail to ensure the preservation of information when mapping classical features to quantum features, thus creating an information gap.

Looking at it from a different angle, if we temporarily set aside quantum theory, Eqn. \eqref{eq:function_v_to_q} reveals that $\mathcal{Q}$ serves as a dimension reduction function, mapping $\mathbb{R}^d$ to $\mathbb{R}^n$ where $n \ll d$. As far as we know, no flawless dimension reduction algorithms can preserve pairwise cosine distances between vectors. Even if a perfect algorithm existed, extending its theory to the quantum realm remains an open question.

\noindent
\textbf{\textit{Problem Visualization}}. Considering the task of face clustering \cite{nguyen2021clusformer}, we assume that a model $\mathcal{M}(x)$ \cite{deng2019arcface} is trained with metric loss functions \cite{wang2018cosface, deng2019arcface} to map a facial image $x$ into a high-dimensional features space. This mapping ensures that similar faces are clustered closely while separating from faces of different identities. As discussed in \cite{nguyen2021clusformer}, recent studies have significantly addressed large-scale clustering challenges within classical machine learning. These methods extensively utilize the discriminative nature of facial features, mainly relying on cosine distance in algorithmic design. However, envisioning a quantum counterpart algorithm that perfectly mirrors these methods reveals a crucial limitation. Despite their potential, quantum algorithms struggle to match the performance of classical ones due to the absence of ideal strategies for encoding classical information into quantum formats, as shown in the Proposition \ref{prop:qig}. 

We illustrate the issue in Fig. \ref{fig:problem_introduce} 
. Specifically, we employ a face recognition model, ResNet50 \cite{he2016deep}, trained with ArcFace \cite{deng2019arcface} on the MSCeleb-1M database \cite{guo2016ms} using classical machine techniques. We randomly select subjects from the hold-out set and extract their facial features. Subsequently, we process the corresponding quantum information of these features according to Eqn. \eqref{eq:full_pipeline}. The boundary between these subjects appears blurred in the quantum machine's perspective, whereas it remains distinct in the classical one. Some samples close together in the classical machine space appear far apart in the quantum space, presenting challenges for quantum algorithms to determine the boundary.

\section{Our Proposed Approach}
\label{sec:proposed_method}
\subsection{Problem Formulation}
Let $x \in \RR^{h \times w \times c}$ denote the input image where $h$, $w$, and $c$ are the image height, width, and number of channels correspondingly. Consider $\mathbf{v} = \mathcal{M}(x)$ is the deep features extracted by a model $\mathcal{M}$. Let $\mathcal{K}$ be the function to measure the gap of information between classical vector $\mathbf{v}$ and its corresponding quantum vector $\mathbf{q}$. Our goal can be presented as in Eqn. \eqref{eq:problem_formulation}.
\begin{equation}
    \label{eq:problem_formulation}
    \textrm{min} \quad \mathcal{K}(\mathbf{v}, \mathbf{q}) = \mathcal{K}(\mathcal{M}(x), \mathcal{Q}(\mathcal{M}(x), \mathcal{E}, \mathcal{O})) \quad \textrm{w.r.t} \quad \mathcal{E} \textrm{,} \mathcal{O} \quad \textrm{and} \quad \mathbf{v} = \mathcal{M}(x)
\end{equation}

\subsection{Quantum Information Preserving Loss}

In Eqn. \eqref{eq:problem_formulation}, only $\mathcal{M}$ and $\mathcal{E}$ are considered trainable. Theoretically, we can optimize either $\mathcal{M}$ or $\mathcal{E}$ to minimize the Eqn. \eqref{eq:problem_formulation}. In this study, however, we concentrate on training $\mathcal{M}$ since, as demonstrated in Eqn. \eqref{eq:full_pipeline}, $\mathbf{q} = \mathcal{M} \circ \mathcal{E} \circ \mathcal{O}$, indicating that $\mathcal{M}$ initiates the quantum encoding process, making it the most critical component to address. 
Let $\mathcal{F}$ represent the task-specific layer to train the feature representation of $x$. $\mathcal{M}$ can be optimized with the objective function as in Eqn. \eqref{eq:objective}.
\begin{equation}
\label{eq:objective}
\theta^*_{\mathcal{M}} = \arg \min_{\theta_{\mathcal{M}}} \mathbb{E}_{x_i \sim p(x_i)} \left[ \mathcal{L} ( \mathcal{F}(\mathcal{M}(x_i)), \hat{y}_i) \right]
\end{equation}
Here, $\hat{y}_i$ and $\mathcal{L}$ denote the ground truth and the loss function, respectively. The common approach (e.g., \cite{deng2009imagenet, he2016deep, liu2022convnet}) typically designs $\mathcal{F}$ as a fully connected layer and employs loss functions such as 
cross-entropy
or metric losses (e.g., \cite{deng2019arcface, wang2018cosface}) for training a classification model. 
For simplicity, we choose 
cross-entropy
as $\mathcal{L}$. It's important to note that, however, $\mathcal{L}$ is also applicable to metric loss functions like ArcFace or CosFace.
\begin{equation}
    \mathcal{L} = - \frac{1}{N} \sum_{i=1}^N \textrm{log} \frac{e^{W_{\hat{y}_i}^\mathsf{T} \mathbf{v}_i + b_j}}{\sum_{j=1}^C e^{W_j^\mathsf{T} \mathbf{v}_i + b_j}}
\label{eq:metric_loss}
\end{equation}
where $W_j \in \RR^d$ denotes the $j^{th}$ column of the weight $W \in \RR^{d \times C}$. $C$ is the number of classes and $b_j \in \RR$ is the bias term. For simply, we fix $b_j = 0$ as in \cite{wang2018cosface}. The equation turns out 
$\mathcal{L} = - \frac{1}{N} \sum_{i=1}^N \textrm{log} \frac{e^{W_{\hat{y}_i}^\mathsf{T} \mathbf{v}_i}}{\sum_{j=1}^C e^{W_j^\mathsf{T} \mathbf{v}_i}}$. 
Interestingly, $W_j$ represents a \textit{center vector} corresponding to class $j$. 
The loss function $\mathcal{L}$ optimizes model $\mathcal{M}$ so that the vector $\mathbf{v}_i$ aligns closely with $W_j$ if they belong to the same class in the feature space. 
Moreover, $W_j^\mathsf{T} \mathbf{v}$ signifies the cosine distance between the two vectors since as in \cite{deng2019arcface,wang2018cosface} these features are normalized, which precisely fulfills the roles of $\ket{\psi_1}$ and $\ket{\psi_2}$ in Proposition \ref{prop:qig}. 
Leveraging this elegant property, we can define $\mathcal{K}$ as the Kullback-Leibler divergence (KL) to minimize the information gap formulated in Eqn. \eqref{eq:problem_formulation} as follows:
\begin{equation}
\begin{split}
    \mathcal{K} &= \frac{1}{N} \sum_{i=1}^N \textrm{KL}\left(W^\mathsf{T} \mathbf{v}_i, S^\mathsf{T} \mathbf{q}_i \right) \\
    &= \frac{1}{N} \sum_{i=1}^N \sum_{j=1}^C \textrm{softmax}(W^\mathsf{T} \mathbf{v}_i)_j 
    \times \textrm{log}\frac{\textrm{softmax}(W^\mathsf{T} \mathbf{v}_i)_j}
    {\textrm{softmax}(S^\mathsf{T} \mathbf{q}_i)_j}
\end{split}
\label{eq:kl_div}
\end{equation}
where $S_j$ is the corresponding quantum information vector of $W_j$ using Eqn. \eqref{eq:function_v_to_q}. In conclusion, we propose a novel loss function named Quantum Information Preserving Loss to train $\mathcal{M}$ as follows: 

\begin{equation}
\label{eq:quantum_information_preserving_loss}
\theta^*_{\mathcal{M}} = 
\arg \min_{\theta_{\mathcal{M}}} \mathbb{E}_{x_i \sim p(x_i)} 
\left[ 
-\textrm{log} \frac{e^{W_{\hat{y}_i}^\mathsf{T} \mathbf{v}_i}}{\sum_{j=1}^C e^{W_j^\mathsf{T} \mathbf{v}_j}} + 
\lambda \times \textrm{KL}\left(W^\mathsf{T} \mathbf{v}_i, S^\mathsf{T} \mathbf{q}_i \right)
\right]
\end{equation}
where $\lambda$ is the loss factor for controlling how much information is preserved. Using this loss function, the model $\mathcal{M}$ can produce the feature $\mathbf{v}$, which is \textit{friendly} with the quantum machine by keeping as much information after the quantum encoding. We also provide the pseudo-code in the Algorithm \ref{algo:pseudo_code}.

\RestyleAlgo{ruled}
\begin{algorithm}[t]
\small
\caption{Pseudo-code for the implementation of Quantum Information Preserving Loss}
\SetKwComment{Comment}{// }{}

\KwData{
\\$\{x_i\}_{i=1}^{N} \in \mathbb{R}^{N \times h \times w \times c}$ : a set of $N$ input images
\\$\{\hat{y}_i\}_{i=1}^{N} \in \mathbb{R}^N$ : a set of $N$ labels
\\$\mathcal{M}$ : feature extractor
\\$\theta_\mathcal{M}$ : trainable parameters of $\mathcal{M}$
\\$\lambda_\mathcal{M}$ : learning rate of $\mathcal{M}$
\\$\lambda$ : loss factor of Quantum Information Preserving loss
}

\While{not convergent} {

    $\mathbf{v}_i \gets \mathcal{M}(x_i)$ \Comment{Extract classical features of the images}
    
    $\mathbf{q}_i \gets \mathcal{Q}(\mathbf{v}_i, \mathcal{E}, \mathcal{O})$ \Comment{Transform into quantum features as Eqn. \eqref{eq:function_v_to_q}}

    $S \gets \mathcal{Q}(W, \mathcal{E}, \mathcal{O})$ \Comment{Transform into quantum center vectors}
    
    $\mathbf{w}_i \gets \textrm{softmax}(W^\top \mathbf{v}_i)$ \Comment{Project classical features into logits}
    
    $\mathbf{u}_i \gets \textrm{softmax}(S^\top \mathbf{q}_i)$ \Comment{Project quantum features into logits}
    
    $\mathcal{L} \gets \frac{1}{N} \sum_{i=1}^N -\log w_{i, \hat{y}_i}$ \Comment{Apply metric loss as Eqn. \eqref{eq:metric_loss}}
    
    $\mathcal{K} \gets \frac{1}{N} \sum_{i=1}^N \sum_{j=1}^C w_{i,j} \log \frac{w_{i,j}}{u_{i,j}}$ \Comment{Apply KL divergence as Eqn. \eqref{eq:kl_div}}

    $\mathcal{L}_\textrm{QIP} \gets \mathcal{L} + \lambda \mathcal{K}$ \Comment{Compute the Quantum Information Preserving Loss}

    $\theta_\mathcal{M} \gets \theta_\mathcal{M} - \lambda_{\mathcal{M}}\nabla_{\theta_\mathcal{M}}\mathcal{L}_\textrm{QIP}$ \Comment{Do backpropagation}
}

\label{algo:pseudo_code}
\end{algorithm}

\section{Experiment Setup and Implementation}

Given that Proposition \ref{prop:qig} implies the information as the relationship between two vectors, i.e., cosine similarity, selecting the model $\mathcal{M}$ optimized for cosine similarity becomes paramount for problem validation and experimental demonstration. Consequently, this study aims for unsupervised clustering tasks, namely face and landmark clustering, as they align well with models trained using cosine-based loss functions. It is important to note that similar problems, such as classification, also apply to our proposed Proposition \ref{prop:qig}. 

\subsection{Experiment Setup}

We follow the experimental framework outlined in previous studies \cite{nguyen2021clusformer,yang2019learning,yang2020learning,shen2023clip,shin2023local,shen2021structure,wang2022ada, nguyen2023fairness}. In essence, our clustering methodology consists of three key stages. First, we train a model $\mathcal{M}(x)$ to extract image features $x$. Second, the $k$ nearest neighbors algorithm, denoted as $\mathbf{K}(x_i, k)$, is utilized to identify the $k$ most similar neighbors of a given sample $x_i$, forming a cluster $\mathbf{\Phi}_i = \mathbf{K}(x_i, k)$. Finally, as clusters $\mathbf{\Phi}_i$ may encompass erroneous samples due to challenges such as database anomalies or imperfect feature representations by $\mathcal{M}$, previous studies have proposed training a model $\mathcal{N}(\mathbf{\Phi}_i)$ to detect and eliminate these inaccuracies, thereby refining the cluster.

In contrast to prior research, we focus on studying this problem from a quantum perspective. It leads to designing modules, namely $\mathcal{M}(x)$ and $\mathcal{N}(\mathbf{\Phi}_i)$, to operate on quantum hardware to the fullest extent possible. While training $\mathcal{M}(x)$ using our proposed methodology constitutes a critical aspect of this study, 
We aim to design $\mathcal{N}(\mathbf{\Phi}_i)$ as a quantum machine learning model, thus enabling the entire pipeline to be executed on a quantum machine as much as possible.

Multiple methodologies have addressed the clustering problem on classical computers. These include traditional techniques \cite{ester1996density, otto2017clustering}, graph-based methodologies \cite{wang2019linkage, yang2020learning, yang2019learning, shen2021structure, shen2023clip, shin2023local}, and transformer-based approaches \cite{nguyen2021clusformer}. While transformer architectures have demonstrated significant success in various computer vision tasks \cite{li2022blip, yu2022coca, zhai2023sigmoid, luo2023lexlip, wang2023equivariant, nguyen2023micron, nguyen2023insect, nguyen2020self, nguyen2019audio,nguyen2019sketch,nguyen2021clusformer,nguyen2022two,nguyen2023algonauts,nguyen2023brainformer,nguyen2023fairness,serna2024video}, their potential in quantum computing remains promising. Adapting the typical transformer architecture for quantum systems, as proposed by \cite{chen2022quantum}, offers added convenience. Although graph-based networks present a possible option, the computational challenge of processing large datasets, such as a (5.2M $\times$ 5.2M) sparse matrix on a quantum machine or even a simulated one, poses limitations. In contrast, transformer models do not encounter such constraints. Hence, inspired by the insights from \cite{nguyen2021clusformer}, we propose redesigning $\mathcal{N}(\mathbf{\Phi}_i)$ as a transformer-based quantum model.

\begin{figure}[t]
    \centering
    \includegraphics[width=0.9\linewidth]{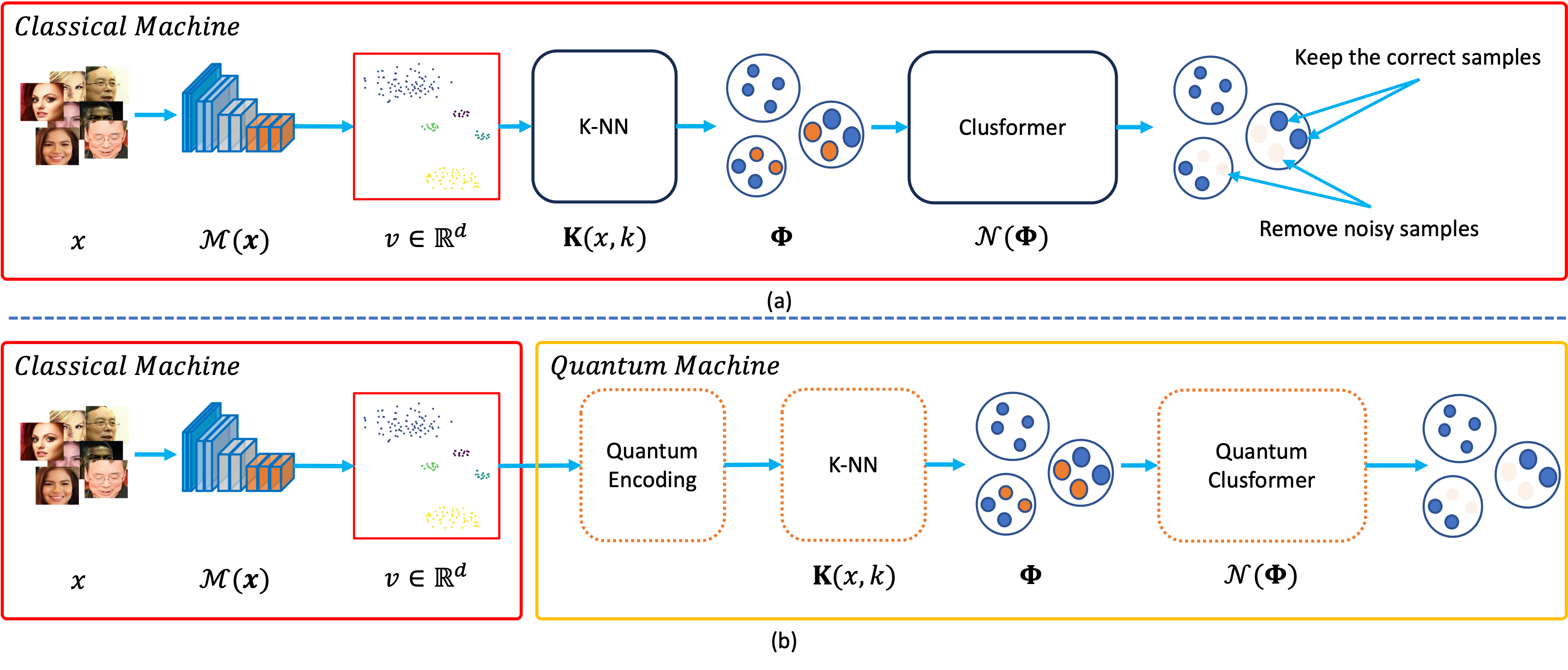}
    \caption{Experiment setup and objective of the clustering problem. Figure (a) depicts the typical experiment setup used by \cite{nguyen2021clusformer} for the classical machine. Figure (b) shows a similar setup. However, only deep model $\mathcal{M}(x)$ retains running on the classical machine, while the rest of the modules are redesigned to run on the quantum computer.}
    \label{fig:exp_setup}
\end{figure}

\subsection{Implementation Details}
We employ ResNet50 architecture to train the model $\mathcal{M}(x)$ as prior works \cite{wang2019linkage,yang2020learning,nguyen2021clusformer}. This model is trained on large-scale datasets like MSCeleb-1M, employing ArcFace \cite{deng2019arcface} for feature representation learning. In addition to ArcFace, we integrate the Quantum Information Preserving Loss outlined in Section \ref{sec:proposed_method} to mitigate information loss during encoding. The loss factor $\lambda$ is configured at 0.5.

To implement the Quantum Clusformer \cite{nguyen2024qclusformer} $\mathcal{N}(\mathbf{\Phi}_i)$, we initially redesign the self-attention layer \cite{vaswani2017attention} tailored for quantum machines. We employ Parameterized Quantum Circuits (PQC) for each Query, Key, and Value layer. We construct transformer blocks suitable for the transformer-based model. Ultimately, we achieve full implementation of the Quantum Clusformer on quantum machines.\footnote{Code will be released upon acceptance}

For the components running on the classical machine, we use the PyTorch framework while we utilize the torchquantum library \cite{hanruiwang2022quantumnas} and cuQuantum to simulate the quantum machine. Since this library relays Pytorch as the backend, we can also leverage GPUs and CUDA to speed up the training process. The models are trained utilizing an 8 $\times$ A100 GPU setup, each with 40GB of memory. The learning rate is initially set to $0.0001$, progressively decreasing to zero following the CosineAnnealing policy \cite{loshchilov2016sgdr}. Each GPU operates with a batch size of $512$. The optimization uses AdamW \cite{loshchilov2017decoupled} for $12$ epochs. Training time for the model $\mathcal{M}$ is approximately 2 hours, and the training time for the Quantum Clusformer $\mathcal{N}(\mathbf{\Phi}_i)$ is about 4 hours. 
\subsection{Datasets and Metrics}

\subsubsection{Datasets}
\begin{figure}[!ht]
    \centering
    \includegraphics[width=0.9\linewidth]{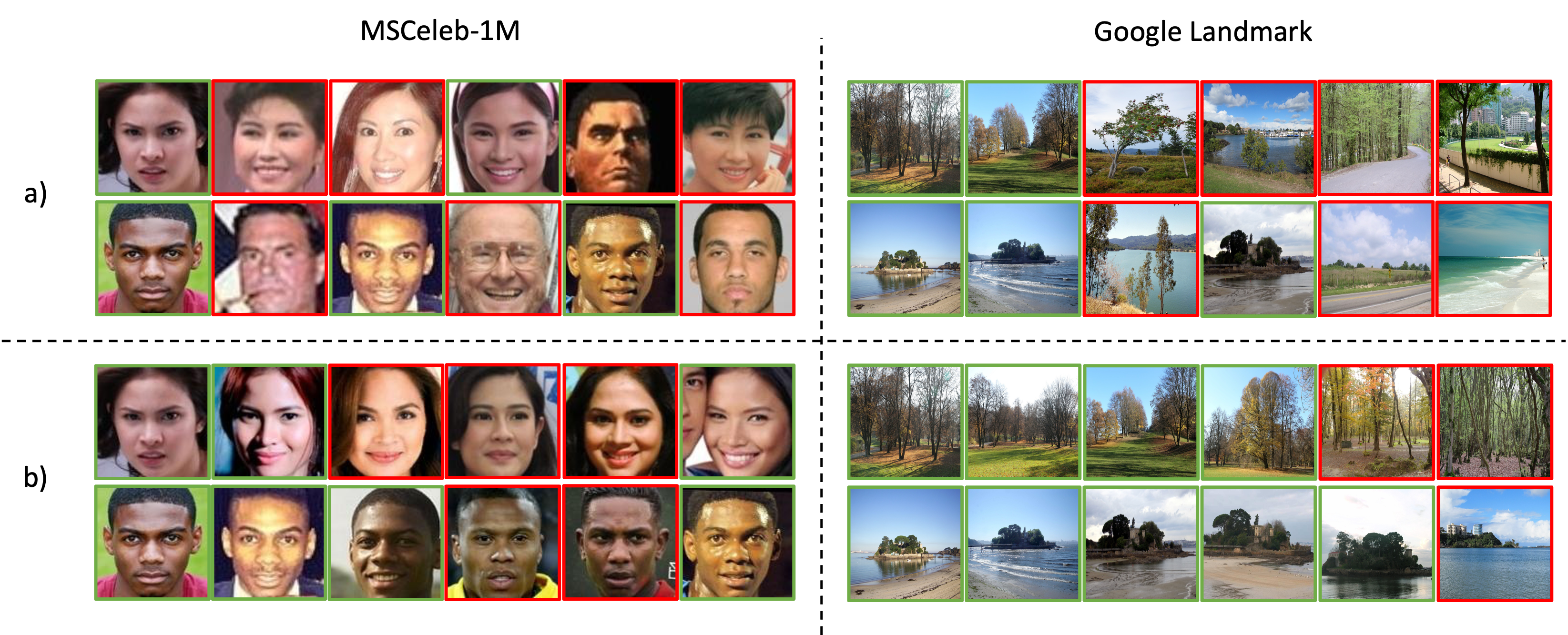}
    \caption{
    The MSCeleb-1M and Google Landmark datasets are illustrated through samples. Each row represents either a subject (for MSCeleb-1M) or a location (for Google Landmark). The first image in each row denotes the center of a cluster $\mathbf{\Phi}_i$, while the subsequent images are the nearest neighbors of the first one, identified through the K-NN algorithm utilizing quantum features. Images bordered in red signify that they belong to a different class than the first image in the row, whereas those bordered in green share the same class as the first image. The clusters obtained without QIP loss in (a) exhibit more noisy samples compared to (b), which are obtained with QIP Loss.
    \textbf{Best view in color.}
    }
    \label{fig:sample_dataset}
\end{figure}

We follow \cite{yang2019learning, yang2020learning} to use MSCeleb-1M \cite{guo2016ms} and \cite{nguyen2021clusformer} to use the Google Landmarks Dataset Version 2 (GLDv2) \cite{weyand2020GLDv2} for experiments. 
\newline
\noindent
\textbf{MSCeleb-1M} \cite{guo2016ms} is a vast face recognition dataset compiled from web sources, encompassing 100,000 identities, with each identity represented by approximately 100 facial images. Nonetheless, the original dataset retains noisy labels. Consequently, we utilize a subset derived from ArcFace \cite{deng2019arcface}, which undergoes improved annotation post-cleaning. This refined dataset comprises 5.8 million images sourced from 85,000 identities. All images undergo pre-processing, involving alignment and cropping to dimensions of $112 \times 112$.
\newline
\noindent
\textbf{The Google Landmarks Dataset Version 2 (GLDv2)} \cite{weyand2020GLDv2} is one of the largest datasets dedicated to visual landmark recognition and identification. Its cleaned iteration comprises 1.4 million images spanning 85,000 landmarks and 800 hours of human annotation. These landmarks span diverse categories and are sourced from various corners of the globe. The dataset exhibits an extremely long-tail distribution, with the number of images per class varying from 0 to 10,000. Compared to face recognition tasks, GLDv2 presents a similar yet notably more challenging scenario. We randomly partition the dataset into three segments, each featuring 28,000 landmarks. Notably, there is no overlap between these partitions. One segment is designated for training the deep visual model and Clusformer, while the remaining segments are reserved for testing purposes. The Fig. \ref{fig:sample_dataset} demonstrates samples from these datasets. 

\subsubsection{Metrics}
To evaluate the approach for the clustering task, we follow \cite{yang2019learning, yang2020learning, nguyen2021clusformer} and use Fowlkes Mallows Score to measure the similarity between two clusters with a set of points.
This score is computed by taking the geometry mean of precision and recall of the point pairs.
Thus, Fowlkes Mallows Score is called Pairwise F-score ($F_P$).
BCubed F-score $F_B$ is another popular metric for clustering evaluation focusing on each data point.

\section{Experimental Results}

\begin{table}[t]
\centering
\caption{Performance on face clustering w.r.t the different number of unlabelled test sets. The terms \textit{Enc} and \textit{Obs} represent quantum encoding strategies and observables, respectively. The \textit{A} and \textit{Z} are amplitude encoding and Pauli-Z observables, correspondingly. QIP is representing for Quantum Information Preserving Loss.}
\addtolength{\tabcolsep}{-5pt}
\fontsize{5.5pt}{7.25pt}\selectfont
\begin{tabular}{c|l|cc|cc|cc|cc|cc|cc}
\midrule
\multirow{2}{*}{{Compute Type}}
& Num. unlabeled & \multicolumn{2}{c}{Setup} & \multicolumn{2}{c}{584K} & \multicolumn{2}{c}{1.74M} & \multicolumn{2}{c}{2.89M} & \multicolumn{2}{c}{4.05M} & \multicolumn{2}{c}{5.21M} \\
\cmidrule{2-14}
& Method / Metrics & Enc & Obs & $F_P$ & $F_B$ & $F_P$ & $F_B$ & $F_P$ & $F_B$ & $F_P$ & $F_B$ & $F_P$ & $F_B$ \\
\midrule
\multirow{15}{*}{Classical}
& K-means \cite{lloyd1982least,sculley2010web}         &- & -& 79.21 & 81.23 & 73.04 & 75.2  & 69.83 & 72.34 & 67.90 & 70.57 & 66.47 & 69.42 \\
& HAC \cite{sibson1973slink}             &- & -& 70.63 & 70.46 & 54.40 & 69.53 & 11.08 & 68.62 & 1.40  & 67.69 & 0.37  & 66.96 \\
& DBSCAN \cite{ester1996density}          &- & -& 67.93 & 67.17 & 63.41 & 66.53 & 52.50 & 66.26 & 45.24 & 44.87 & 44.94 & 44.74 \\
& ARO \cite{otto2017clustering}             &- & -& 13.60 & 17.00 & 8.78  & 12.42 & 7.30  & 10.96 & 6.86  & 10.50 & 6.35  & 10.01 \\
& CDP \cite{zhan2018consensus}             &- & -& 75.02 & 78.70 & 70.75 & 75.82 & 69.51 & 74.58 & 68.62 & 73.62 & 68.06 & 72.92 \\
& L-GCN \cite{wang2019linkage}           &- & -& 78.68 & 84.37 & 75.83 & 81.61 & 74.29 & 80.11 & 73.70 & 79.33 & 72.99 & 78.60 \\
& LTC \cite{yang2019learning}             &- & -& 85.66 & 85.52 & 82.41 & 83.01 & 80.32 & 81.10 & 78.98 & 79.84 & 77.87 & 78.86 \\
& GCN-V \cite{yang2020learning}           &- & -& 87.14 & 85.82 & 83.49 & 82.63 & 81.51 & 81.05 & 79.97 & 79.92 & 78.77 & 79.09 \\
& GCN-VE \cite{yang2020learning}          &- & -& 87.93 & 86.09 & 84.04 & 82.84 & 82.10 & 81.24 & 80.45 & 80.09 & 79.30 & 79.25 \\
& Clusformer \cite{nguyen2021clusformer}      &- & -& 88.20 & 87.17 & 84.60 & 84.05 & 82.79 & 82.30 & 81.03 & 80.51 & 79.91 & 79.95 \\
& Pair-Cls \cite{liu2021learn} &- & -& 90.67 & 89.54 & 86.91 & 86.25 & 85.06 & 84.55 & 83.51 & 83.49 & 82.41 & 82.40 \\
& STAR-FC \cite{shen2021structure}       &- & -& 91.97 & - & 88.28 & 86.26 & 86.17 & 84.13 & 84.70 & 82.63 & 83.46 & 81.47 \\
& Ada-NETS \cite{wang2022ada}            &- & -& 92.79 & 91.40 & 89.33 & 87.98 & 87.50 & 86.03 & 85.40 & 84.48 & 83.99 & 83.28 \\
& LCE-PCENet \cite{shin2023local}        &- & -& 94.64 & 93.36 & 91.90 & 90.78 & 90.27 & 89.28 & 88.69 & 88.15 & 87.35 & 87.28 \\
& CLIP-Cluster \cite{shen2023clip}       &- & -& - & - & 91.44 & 89.44 & 89.95 & 87.75 & 88.93 & 86.78 & 87.99 & 85.85 \\
& $\textrm{Clusformer}^{\dagger}$       &- & - & 86.49 &	89.82 &	84.40 &	87.84 &	82.41 &	85.86 &	80.42 &	83.87 &	78.33 &	81.73 \\
\midrule
\multirow{2}{*}{Quantum}
& QClusformer             &A & Z & 83.68 &	86.89 &	81.93 &	85.19 &	79.77 &	83.05 &	78.32 &	81.41 &	76.15 &	79.29 \\
& QClusformer  + QIP Loss &A & Z & \textbf{87.18} &	\textbf{91.01} &	\textbf{85.14} &	\textbf{89.32} &	\textbf{83.19} &	\textbf{87.34} &	\textbf{81.59} &	\textbf{85.83} &	\textbf{79.40} &	\textbf{83.78} \\

\midrule
\end{tabular}
\par
\label{tab:ms1m_results}
\end{table}

\subsection{Performance on MSCeleb-1M Clustering}
The performance of our proposed method is shown in the Table \ref{tab:ms1m_results}.
To begin, we define QClusformer as the Clusformer operating on a quantum machine for ease of reference. However, due to hardware constraints, we can only emulate QClusformer with fewer layers/transformer blocks than the original model \cite{nguyen2021clusformer}. To ensure a fair evaluation, we initially retrain the Clusformer, denoted as $\textrm{Clusformer}^{\dagger}$, on a classical machine using identical configurations to those of QClusformer, explicitly setting the number of encoders to 1. The training process is outlined in Fig. \ref{fig:exp_setup}(a). As a result, the performance of $\textrm{Clusformer}^{\dagger}$ is slightly inferior to the original model. Notably, the $F_P$ metric decreases from 88.20\% to 86.49\% on the 584K test set, representing an approximate 2\% reduction. It consistently maintains marginally lower performance across both $F_B$ and $F_P$ on the remaining test sets.

\noindent
Then, we train QClusformer with the strategy as in Fig. \ref{fig:exp_setup}(b). Our chosen encoding strategy is amplitude, paired with Pauli-Z as the observable for the baseline. There is a notable decline in performance, approximately 2.8\%. %
However, employing the QIP Loss function within the same setup is a potent remedy for bridging the information gap between quantum and classical features, resulting in a notable performance recovery. Noted that QClusformer with QIP Loss achieves 87.18\% and 91.01\% on $F_P$ and $F_B$, respectively, on the 584K test set, surpassing $\textrm{Clusformer}^{\dagger}$ by 0.6\% and 3.2\%, respectively. Similar trends are observed across all test sets of MSCeleb-1M.

\noindent
These findings underscore the competitive performance of Quantum Clusformer, particularly when leveraging with QIP Loss. Notably, its performance surpasses that of the best-performing Clusformer with a complete setup on a classical machine, signaling the promising capabilities of quantum computing in the clustering problem.

\subsection{Performance on Google-Landmark Clustering}
This section compares the proposed method's performance on the Google Landmark dataset, a visual landmark clustering dataset shown in Table \ref{tab:abl_landmark_results}. 
The experimental setups and evaluation protocols are similar to the previous MSCeleb-1M section and in the prior work, \cite{nguyen2021clusformer}. 
Similar results to those obtained with the MSCeleb-1M database are observed. 
Specifically, $\textrm{Clusformer}^{\dagger}$, when runs on a classical machine, achieves 17.74\% and 38.80\% in terms of $F_P$ and $F_B$ respectively. 
However, when the model operates on a quantum machine named QClusformer, its performance drops significantly to 13.20\% and 35.63\% for $F_P$ and $F_B$, respectively. 
Nonetheless, by using the QIP Loss function, the performance rebounds to 19.02\% for $F_P$ and 40.28\% for $F_B$, surpassing that of $\textrm{Clusformer}^{\dagger}$ and remaining competitive with the original Clusformer which has 19.32\% and 40.63\% of $F_P$ and $F_B$.

\begin{table}[!ht]
\centering
\caption{Performance on landmark clustering w.r.t different quantum encoding and observables.}
\begin{tabular}{c|l|cc|cc}
\midrule
Compute Type & Methods           & Enc & Obs & $F_P$    & $F_B$    \\
\midrule
\multirow{10}{*}{Classical} 
& K-means \cite{lloyd1982least,sculley2010web}      & - & - & 8.52  & 14.02 \\
& HAC \cite{sibson1973slink}          & - & - & 0.2   & 20.88 \\
& DBSCAN \cite{ester1996density}       & - & - & 0.97  & 17.38 \\
& Spectral \cite{ho2003clustering}     & - & - & 6.93  & 18.28 \\
& ARO \cite{otto2017clustering}          & - & - & 0.32  & 10.54 \\
& L-GCN \cite{wang2019linkage}        & - & - & 14.08 & 36.35 \\
& GCN-V \cite{yang2020learning}       & - & - & 16.1  & 34.86 \\
& GCN-VE \cite{yang2020learning}     & - & - & 10.2  & 30.23 \\
& Clusformer \cite{nguyen2021clusformer} & - & - & 19.32 & 40.63 \\
& $\textrm{Clusformer}^{\dagger}$           & - & - & 17.74 & 38.80 \\
\midrule
\multirow{8}{*}{Quantum}
& QClusformer                               & A & Z & 13.20 & 35.63 \\
& QClusformer  + QIP Loss                   & A & Z & \textbf{19.02} & \textbf{40.28} \\ 
\cmidrule{2-6}
& QClusformer  + QIP Loss                   & A & X & 18.50 &	38.86 \\
& QClusformer  + QIP Loss                   & A & XZ & 17.58 &	37.84 \\
\cmidrule{2-6}
& QClusformer  + QIP Loss                   & P & Z & 17.02 &	36.50 \\
& QClusformer  + QIP Loss                   & $\textrm{U}_3$ & Z & 16.64 &	36.04 \\  
& QClusformer  + QIP Loss                   & $\textrm{U}_3$ & Y & 16.41 &	36.68 \\
\midrule
\end{tabular}
\label{tab:abl_landmark_results}
\end{table}

\subsection{Ablation Studies}
This ablation study section practically proves the Proposition \ref{prop:qig}. 

\noindent
\textbf{QIP Works With Different Encoding Strategies}.
In Proposition \ref{prop:qig}, we present the information gap between quantum and classical machines across various encoding strategies. To demonstrate the efficiency of our proposed method with diverse encoding approaches, we initially hold observables constant, specifically the Pauli-Z, and subsequently change between phase and $U_3$ encoding \cite{benedetti2019parameterized}. Unlike amplitude and phase encoding, $U_3$ represents a Parameterized Quantum Circuit (PQC) with trainable parameters. The performances of these configurations are detailed in Table \ref{tab:abl_encoding}. Remarkably, the QClusformer, trained with QIP Loss, the Pauli-Z observable, and either phase or $U_3$ encoding strategies, consistently outperforms the standalone QClusformer. It underscores the adaptability of the QIP Loss across diverse encoding strategies. Notably, phase and $U_3$ encoding show inferior performance compared to amplitude. As we mentioned in the previous section, the amplitude is naturally fit for the clustering problem than other strategies.

\begin{table}[!ht]
\centering
\caption{Ablation studies on different encoding strategies of the MSCeleb-1M.}
\addtolength{\tabcolsep}{-5pt}
\fontsize{7.5pt}{9.25pt}\selectfont
\begin{tabular}{c|l|cc|cc|cc|cc|cc|cc}
\midrule
& Num. unlabeled & \multicolumn{2}{c}{Setup} & \multicolumn{2}{c}{584K} & \multicolumn{2}{c}{1.74M} & \multicolumn{2}{c}{2.89M} & \multicolumn{2}{c}{4.05M} & \multicolumn{2}{c}{5.21M} \\
\cmidrule{2-14}
& Method / Metrics & Enc & Obs & $F_P$ & $F_B$ & $F_P$ & $F_B$ & $F_P$ & $F_B$ & $F_P$ & $F_B$ & $F_P$ & $F_B$ \\
\midrule
\multirow{2}{*}{\rotatebox[origin=c]{90}{Cls}}
& Clusformer \cite{nguyen2021clusformer}      &- & -& 88.20 & 87.17 & 84.60 & 84.05 & 82.79 & 82.30 & 81.03 & 80.51 & 79.91 & 79.95 \\
& $\textrm{Clusformer}^{\dagger}$       &- & - & 86.49 &	89.82 &	84.40 &	87.84 &	82.41 &	85.86 &	80.42 &	83.87 &	78.33 &	81.73 \\
\midrule
\multirow{4}{*}{\rotatebox[origin=c]{90}{Quantum}}
& QClusformer             &A & Z & 83.68 &	86.89 &	81.93 &	85.19 &	79.77 &	83.05 &	78.32 &	81.41 &	76.15 &	79.29 \\
& QClusformer  + QIP Loss &A & Z & \textbf{87.18} &	\textbf{91.01} &	\textbf{85.14} &	\textbf{89.32} &	\textbf{83.19} &	\textbf{87.34} &	\textbf{81.59} &	\textbf{85.83} &	\textbf{79.40 }&	\textbf{83.78} \\
& QClusformer  + QIP Loss &P & Z  & 86.42 &	88.41 &	84.73 &	86.60 &	82.82 &	84.62 &	81.36 &	83.06 &	79.33 &	81.01 \\
& QClusformer  + QIP Loss &$\textrm{U}_3$ & Z  & 85.20 &	89.64 &	83.68 &	87.71 &	81.49 &	85.62 &	80.03 &	83.73 &	77.84 &	81.79 \\
\midrule
\end{tabular}
\label{tab:abl_encoding}
\end{table}

\noindent
\textbf{QIP Works With Different Observables}. The intuition of these ablation studies is similar to the encoding above strategies. In particular, we fix the encoding strategies as amplitude while experimenting with various observables, i.e., $Z$, $X$, and $XZ$ (a combination of measuring both $X$ and $Z$ coordinates). As depicted in Table \ref{tab:abl_observable}, QClusformer exhibits the highest accuracy in $F_P$ and $F_B$ when utilizing the $Z$ observable, while both $X$ and $XZ$ show slight decreases. When dealing with the Pauli-Y observable, amplitude strategies prove ineffective as they result in all-zero measurements. Consequently, we select $U_3$ for encoding and compare the performance of Pauli-Y versus Pauli-Z. Interestingly, the performance using Pauli-Y remains relatively unchanged compared to Pauli-Z. Nonetheless, these configurations still significantly outperform QClusformer alone, underscoring the versatility of the Quantum Information Processing (QIP) approach across diverse observables.

\begin{table}[!ht]
\centering
\caption{Ablation studies on different observables of MSCeleb-1M.}
\addtolength{\tabcolsep}{-5pt}
\fontsize{7.5pt}{9.25pt}\selectfont
\begin{tabular}{c|l|cc|cc|cc|cc|cc|cc}
\midrule
& Num. unlabeled & \multicolumn{2}{c}{Setup} & \multicolumn{2}{c}{584K} & \multicolumn{2}{c}{1.74M} & \multicolumn{2}{c}{2.89M} & \multicolumn{2}{c}{4.05M} & \multicolumn{2}{c}{5.21M} \\
\cmidrule{2-14}
& Method / Metrics & Enc & Obs & $F_P$ & $F_B$ & $F_P$ & $F_B$ & $F_P$ & $F_B$ & $F_P$ & $F_B$ & $F_P$ & $F_B$ \\
\midrule
\multirow{2}{*}{\rotatebox[origin=c]{90}{Cls}}
& Clusformer \cite{nguyen2021clusformer}      &- & -& 88.20 & 87.17 & 84.60 & 84.05 & 82.79 & 82.30 & 81.03 & 80.51 & 79.91 & 79.95 \\
& $\textrm{Clusformer}^{\dagger}$       &- & - & 86.49 &	89.82 &	84.40 &	87.84 &	82.41 &	85.86 &	80.42 &	83.87 &	78.33 &	81.73 \\
\midrule
\multirow{6}{*}{\rotatebox[origin=c]{90}{Quantum}}
& QClusformer             &A & Z & 83.68 &	86.89 &	81.93 &	85.19 &	79.77 &	83.05 &	78.32 &	81.41 &	76.15 &	79.29 \\
& QClusformer  + QIP Loss &A & Z & \textbf{87.18} &	\textbf{91.01} &	\textbf{85.14} &	\textbf{89.32} &	\textbf{83.19} &	\textbf{87.34} &	\textbf{81.59} &	\textbf{85.83} &	\textbf{79.40} &	\textbf{83.78} \\
\cmidrule{2-14}
& QClusformer  + QIP Loss &A & X  & 86.40 &	90.30 &	84.32 &	88.76 &	82.40 &	86.85 &	80.95 &	85.27 &	79.03 &	83.29 \\
& QClusformer  + QIP Loss &A & XZ & 86.74 &	89.28 &	84.79 &	87.23 &	82.58 &	85.02 &	81.17 &	83.49 &	79.03 &	81.50 \\
\cmidrule{2-14}
& QClusformer  + QIP Loss &$\textrm{U}_3$ & Z  & 85.20 &	89.64 &	83.68 &	87.71 &	81.49 &	85.62 &	80.03 &	83.73 &	77.84 &	81.79 \\
& QClusformer  + QIP Loss &$\textrm{U}_3$ & Y  & 84.35 &	90.54 &	82.45 &	88.96 &	80.27 &	86.97 &	78.50 &	85.06 &	76.46 &	83.07 \\
\midrule
\end{tabular}
\vspace{-4mm}
\label{tab:abl_observable}
\end{table}

\noindent
\textbf{The role of $\lambda$ - QIP Loss Factor}. We investigate the impact of the control factor $\lambda$ for managing QIP Loss on the performance. To achieve this, we conduct experiments using a subset of 584K samples from the MSCeleb-1M dataset. The experimental configurations remain consistent with those outlined in the previous section, i.e., employing amplitude encoding and Pauli-Z observable.

\begin{figure}[t]
\begin{minipage}[b]{0.34\linewidth}\centering
    \includegraphics[width=1.0\linewidth]{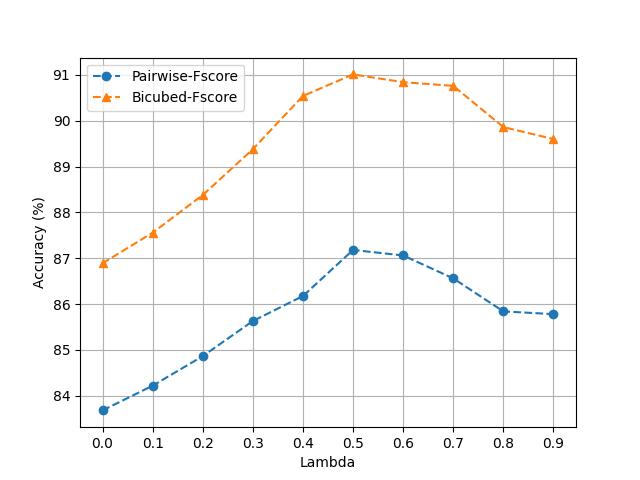}
    \caption{Ablation studies on different QIP Loss factor $\lambda$}
    \label{fig:abl_lambda}
\end{minipage}
\hfill
\begin{minipage}[b]{0.64\linewidth}\centering
    \includegraphics[width=1.0\linewidth]{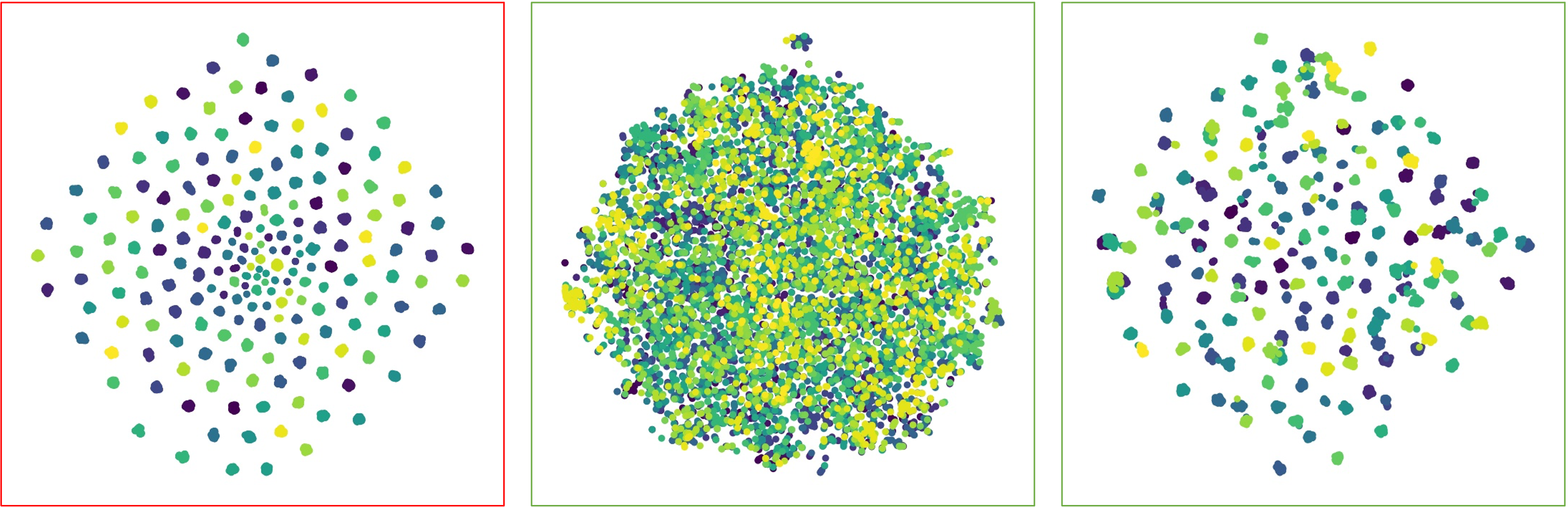}

    \caption{Ablation studies on features representation using QIP Loss. From left to right, the first image presents classical features, the second one presents quantum features w/o QIP Loss, and the last one shows the quantum features optimized by QIP Loss.}
    \label{fig:abl_feature_representation}
\end{minipage}
\end{figure}

The results are shown in Fig. \ref{fig:abl_lambda}. When $\lambda = 0$, indicating the absence of QIP Loss utilization, the performance stands at 83.68\% and 86.89\% for $F_P$ and $F_B$ respectively, as detailed in Table \ref{tab:ms1m_results} above. Gradually increasing this parameter yields a steady enhancement in performance. However, the peak performance is attained at $\lambda=0.5$, after which a decline is observed. This phenomenon is due to the role of QIP Loss in minimizing the disparity between quantum and classical features. According to Proposition \ref{prop:qig}, the gap towards zero only when two vectors $\textbf{v}_1$ and $\textbf{v}_2$ are identical. In this case, the model $\mathcal{M}$ generates similar features irrespective of input images, leading to model collapse and failure in distinguishing samples from distinct classes. Hence, it is necessary to control $\lambda$ to prevent such collapse. Our investigation found that the optimal value for $\lambda$ within this framework is 0.5.

\noindent
\textbf{Quantum Feature Representations}. We investigate how QIP Loss helps to align the features in the quantum computer as in Fig. \ref{fig:abl_feature_representation}. We randomly select 200 subjects from 581K part of MSCeleb-1M to extract the features. We employ T-SNE to reduce the dimension from 256 to 2 and visualize these features in the 2D space. From left to right, the first image (with a red border) indicates the classical features. The second image (with a green border) illustrates the quantum features of these subjects without training with QIP Loss, and the last one demonstrates the quantum features optimized by QIP Loss.

\noindent
\textbf{Performance of feature extractor - $\mathcal{M}$.} \Rone{Since $\mathcal{M}$ is trained by a combination of ArcFace \cite{deng2019arcface} and our proposed QIP Loss, it is important to evaluate the effectiveness of $\mathcal{M}$ and verify how QIP Loss affects to its performance. We follow the same evaluation protocol as in \cite{deng2019arcface}. In particular, we evaluate the face verification accuracy of $\mathcal{M}$ on the IJBC \cite{ijbc} database. The results are reported in the Table \ref{tab:efficientness_m}. As the baseline, the performance of Resnet50 without using QIP Loss on IJBC is 96.140\%. We observe a slight drop to 96.068 when incorporating QIP Loss with the factor by $\lambda = 0.5$. However, the lambda is increased to $\lambda = 0.9$, the performance is reduced by 4\% approximately. The reason for that drop can be explained in the section above where the feature representation tends to collapse when increasing $\lambda$. } 

\begin{table}
    \centering
    \caption{Face verification accuracy of feature extractor $\mathcal{M}$ on IJBC database.}
    \begin{tabular}{c|l|l|c}
        Method & Loss & $\lambda$ & Accuracy (\%) \\
        \hline
        Resnet50 & ArcFace & 0 & 96.140 \\
        Resnet50 & ArcFace + QIP Loss & 0.5 & 96.068 \\
        Resnet50 & ArcFace + QIP Loss & 0.9 & 92.382 \\
        \hline
    \end{tabular}
    \label{tab:efficientness_m}
\end{table}

\noindent
\textbf{Comparison with classical method.} \Rtwo{
Since the problem can be treated as a representation learning task, we compare our method to a classical machine learning approach in this section. Specifically, we choose the Support Vector Machine (SVM), a kernel-based feature representation method, for the comparison. Following \cite{schuld2021supervised}, we implement a Quantum SVM algorithm that can be executed on a quantum computer. This algorithm comprises two main components: quantum encoding and measurement, i.e., Parameterized Quantum Circuit (PQC). Unlike the aforementioned training strategy, we do not train $\mathcal{M}$ jointly with Quantum SVM. Instead, we train the Quantum SVM separately, using classical features $v$ as input to perform a classification task. After training, the corresponding quantum features are utilized to train the Quantum Clusformer $\mathcal{N}(\mathbf{\Phi}_i)$. The performance results are presented in Table \ref{tab:compare_quantumsvm}. Using Quantum SVM for quantum feature representation results in a significant performance drop. It achieves $F_P$ and $F_B$ scores of 80.3\% and 82.82\%, respectively, which is about 7\% lower than our proposed method approximately. This decline in performance is because Quantum SVM is designed for a \textit{close-set} problem, whereas unsupervised clustering addresses an \textit{open-set} problem. While Quantum SVM may provide a good quantum feature representation for the training set, it struggles with the testing set, leading to poor feature distinction and, consequently, the worst performance.}

\begin{table}[!hb]
    \centering
    \caption{Performance comparison with classical method Quantum SVM on 584K subject of MSCeleb-1M}
    \begin{tabular}{l|l|lc|c|c}
        Env & Method & Enc & Obs & $F_P$ & $F_B$ \\
        \hline
        Cls & Clusformer & - & - & 88.29 & 87.17 \\
        Cls & Clusformer & - & - & 86.49 & 89.82 \\
        Quantum & QClusformer & A & Z & 83.68 & 86.89 \\
        Quantum & QClusformer + QIP Loss & A & Z & 87.18 & 91.01 \\
        Quantum & Quantum SVM & A & Z & 80.30 & 82.82 \\
        \hline
    \end{tabular}
    \label{tab:compare_quantumsvm}
\end{table}

\section{Conclusion}
This paper revisits the quantum visual feature encoding strategies employed in quantum machine learning with computer vision applications. We identify a significant Quantum Information Gap (QIG) issue stemming from current encoding methods, resulting in non-discriminative feature representations in the quantum space, thereby challenging quantum machine learning algorithms. To tackle this challenge, we propose a simple yet effective solution called Quantum Information Preserving Loss. Through empirical experiments conducted on various large-scale datasets, we demonstrate the effectiveness of our approach, achieving state-of-the-art performance in clustering problems on quantum machines. Our insights into quantum encoding strategies are poised to stimulate further research efforts in this domain, prompting researchers to focus on designing more effective quantum machine learning algorithms.

\section{Discussion}
Since quantum machines have limited access to the general public, the experiments were carried out through noise-free simulation systems such as torchquantum and cuQuantum. However, real-world scenarios may involve noise within the system, leading to uncertain quantum state measurements and affecting overall performance. Despite this limitation, the theoretical problem of QIG persists. It is crucial to figure out that quantum machine learning algorithms must confront these dual challenges of QIP and noise. We anticipate that addressing these issues will attract significant research attention in future endeavors.

\backmatter

\bmhead{Acknowledgements}
This work is partly supported by MonArk NSF Quantum Foundry, supported by the National Science Foundation Q-AMASE-i program under NSF award No. DMR-1906383. It acknowledges the Arkansas High-Performance Computing Center for providing GPUs.

\section*{Declarations}

\begin{itemize}
\item Data availability: The MSCeleb-1M \cite{guo2016ms} is no longer available due to ethical and privacy concerns. The Google-Landmark database \cite{weyand2020GLDv2} is public available at \href{https://github.com/cvdfoundation/google-landmark
}{https://github.com/cvdfoundation/google-landmark}.
\item Code availability: The code will be available upon acceptance.
\item Author contribution: X.B wrote the main manuscript. H.Q prepared pseudo code, result tables, and experiment setups. H.C and S.K provided fundamental materials of the quantum machine. K.L discussed the novelty and the research direction. All the authors revised the manuscript. 
\end{itemize}

\bibliography{sn-bibliography}%

\end{document}